# Regularization of a mesh generated with the Mesh-Matching algorithm: Application to exophtalmia and maxillofacial computer aided surgery

Luboz Vincent*, Swider Pascal**, Payan Yohan*

\* TIMC Laboratory, UMR CNRS 5525, University J. Fourier, 38706 La Tronche, France
\*\* Biomechanics Laboratory, IFR 30, Purpan University Hospital, 31059 Toulouse cedex 3, France

Corresponding authors:
Payan Yohan, Luboz Vincent
Laboratoire TIMC/IMAG,
UMR CNRS 5525,
Institut d'Ingénierie de l'Information de Santé
Pavillon Taillefer – Faculté de Médecine
38706 La Tronche
France

Tel: +33 4 56 52 00 01
 Fax: +33 4 56 52 00 55
 e-mail: yohan.payan@imag.fr, vincent.luboz@imag.fr

# ABSTRACT

*Objective*:

An automatic mesh regularization procedure was proposed in order to achieve the numerical feasibility of the Finite Element Analysis.

*Design*:

The algorithm has been implemented for tetrahedrons, wedges and hexahedrons.

*Background*:

One of the main drawbacks of three-dimensional model generation is consumption due to the manual three-dimensional meshing procedure. In a previous study, the authors demonstrated the ability of the Mesh Matching algorithm to automatically generate customized three-dimensional meshes from an already existing model. For anatomical structures, some element irregularities can occur after the use of the Mesh-Matching algorithm, making any finite element analysis impossible.

*Methods*:

A process based on the study of the singularity of the Jacobian matrix is used to iteratively correct them.

*Results*:

The method was successfully evaluated on an academic test case (cubic structure meshed with hexahedrons) and on clinical applications (face and orbit meshes).

<u>*Conclusions*</u>:

The use of the combination of the Mesh-Matching algorithm with the regularization phase presented here seems to automatically generate new finite element meshes. Nevertheless, no guaranty, in terms of convergence, can be given, since the regularization algorithm is iterative.

**Relevance:**

To our knowledge, this is the first time that an algorithm proposes to automatically generate patient-specific finite element meshes from an existing generic finite element mesh.

## 1. Introduction

Finite Element (FE) analysis is a widely used method in the field of biomechanics and customized meshes are of great interest since they can integrate both geometry and mechanical properties of the patient. Recently, the Mesh Matching (M-M) algorithm (Couteau et al., 2000) was introduced to automatically generate customised hexahedron and wedge 3D patient meshes from an existing 3D generic mesh. The algorithm was successfully applied to proximal (Couteau et al, 2000) and entire (Luboz et al, 2001) femora. However, the application to a more complicated geometry, namely a FE model of the human face (Chabanas and Payan, 2000), provided non-satisfying mesh irregularities that made the mechanical analysis impossible.

In commercial products, automatically meshing a 3D structure generally uses the tetrahedral meshing technique, which is the most common form of unstructured mesh generation. This technique is frequently based on the Delaunay criterion (Delaunay et al., 1934) followed by the advancing front technique (Lo, 1991). The advantage of hexahedral meshes, compared with tetrahedral meshes, is their increased accuracy. Their drawback is that hexahedral meshing of complicated geometry is difficult (Owen, 1998) and requires a large amount of manual intervention.

Before numerical computation can be carried out, the manually designed meshes often need to be corrected, which is also time consuming. Several regularization techniques are proposed in the literature and are generally adapted to tetrahedral elements. They involve a reconnection algorithm (Joe, 1995) or a node point adjustment method (Amezua et al., 1995). Geometrically correcting a set of elements inside a 3D mesh is a complex problem without any straightforward solution (Cannan et al., 1993; Freitag and Plassmann, 1999). Indeed, correcting a single element can distort its neighbours although they were originally regular. Elements must therefore be considered together for the mesh to be corrected.

The goal of this study is to develop an automatic mesh regularization procedure in order to achieve the numerical feasibility of the Finite Element Analysis. The method can be applied to any element type (tetrahedron, hexahedron, wedge). The locations of the nodes of irregular elements are iteratively corrected using the Jacobian determinant variations. The method is first evaluated on an academic test case (cubic structure meshed with hexahedrons). The method is then applied to seven human faces to investigate the feasibility of a clinical application.

**2. Materials and Methods**

The patient mesh generation is obtained in two steps. First, the M-M algorithm (Couteau et al., 2000) is applied to a standard model. Then, irregular elements that might have been generated by the M-M algorithm are automatically regularized. This paper focuses on the second phase and will only briefly describe the M-M algorithm.

2.1. M-M Algorithm Application

The steps of the M-M algorithm:

1. A FE model of the structure is chosen. This model is often built from a standard patient morphology. Its 3D mesh is assumed to be optimal in terms of mesh refinement and mesh regularity. This model is called the "generic model" since it is used in the M-M algorithm as a starting point to define other FE meshes of the same anatomical structure corresponding to other patient morphologies.
2. The external surface of the patient anatomical structure is extracted through CT (or MRI) acquisition. On each CT (or MRI) slice, the external contour of the structure is segmented, providing a set of 3D points located on the surface.

3. An elastic registration method, originally proposed in the field of computer-assisted surgery (Lavallée et al., 1995; 1996), is used to match the extracted patient surface points with the nodes located on the external surface of the generic FE model. This matching aims at finding a volumetric transform **T**, which is a combination of global (rigid) and local (elastic) transforms. The idea underlying the matching algorithm consists (1) in aligning the two datasets (the rigid part of **T**) and (2) in finding local cubic B-Splines functions (Szeliski and Lavallée, 1996). The unknowns of the transform are all the B-Splines parameters. Those parameters are obtained through an optimization process (the Levenberg-Marquardt algorithm and a modified conjugate gradient algorithm) that aims at minimizing the distance between the two surfaces, namely the points extracted from the patient data and the external nodes of the generic FE model.

4. The volumetric transform **T** is then applied to every node of the FE generic mesh, namely the nodes located on the external surface as well as the internal nodes that define the FE volume. A new volumetric mesh is thus automatically obtained by assembling the transformed nodes into elements, with a topology similar to that of the generic FE model: same number of elements and same element types.

### 2.2. Regularization of the Mesh

#### 2.2.1 Regularity Criteria

Before improving the quality of the Finite Element mesh, the regularization phase checks whether each element of the mesh is regular. This regularity notion is associated with the Jacobian matrix transform, coupling the reference element (unit reference framework) and the actual element (real reference framework) (Touzot and Dhatt, 1984, Zienkiewicz and Taylor, 1994).

Finite Element Analysis is carried out only if the transform can be computed on each point inside the element, that is to say if the Jacobian determinant value (det**J**) is larger than zero anywhere inside the element. The Jacobian determinant det**J** is computed at each node of each element. If a negative or nil value is obtained for one of the nodes, the element is classified as irregular.

### 2.2.2 Regularization Algorithm

The regularization algorithm consists of an iterative process: nodes of irregular elements are slightly shifted at each step, until each element becomes regular. In the following development, the subscript variables are: $k$ - irregular element; $i$ - node(s) of element $k$ with nil or negative Jacobian determinant; $j$ - nodes attached to element $k$; $n$ - number of nodes of element $k$.

The regularization procedure consists of two main steps:
- Computation of the Jacobian determinant (which has no dimension) at each node of the mesh and detection of irregular element $k$ (det$\mathbf{J_i} \leq 0$).
- Automatic correction of irregular element $k$ using a numerical sensitivity procedure based on gradient evaluation.

The idea is to iteratively move each node $i$ (where det$\mathbf{J_i} \leq 0$) in a direction that tends to increase the det$\mathbf{J_i}$ value. As an analytical expression of the gradient vector $\nabla(\text{det}\mathbf{J_i})_j$ can be found. The algorithm consists of moving the node in the direction of the gradient vector in order to increase det$\mathbf{J_i}$.

As expressed in equation (1), the gradient vector $\nabla(\text{det}\mathbf{J_i})_j$ (whose dimension is : length$^{-1}$) is first computed using actual coordinates $\mathbf{X}_j(x_j, y_j, z_j)$ of nodes $j$ attached to the distorted element *k* (with a first order Taylor Series). This gradient vector provides an evaluation of the sensitivity of the geometrical transform (reference framework / actual framework) to the nodes locations. Analytical expressions of $\det\mathbf{J_i}$ and $\nabla(\det\mathbf{J_i})_j$ are derived using a computer algebra system (Maple©).

$$\nabla(\det\mathbf{J_i})_j = \begin{bmatrix} \frac{\partial \det\mathbf{J_i}}{\partial x_i}(x_j) \\ \frac{\partial \det\mathbf{J_i}}{\partial y_i}(y_j) \\ \frac{\partial \det\mathbf{J_i}}{\partial z_i}(z_j) \end{bmatrix} \qquad \text{where } j = 1...n. \tag{1}$$

The directional vector $\mathbf{V}_j$, expressed by equation (2), is determined for updating the node locations. The dimension of $\mathbf{V}_j$ and its Euclidian norm $\|\mathbf{V}_j\|$ is length. For a node with index *j*, the gradient vectors (1) are summed at the element *k* level. If *n* is the number of nodes of this element *k*, the gradient vector is computed and summed for each node *i* (from 1 to *n*) of the element. Taking into account that only gradient vectors of irregular nodes are summed, a coefficient $\alpha_i$ is introduced. The value of this coefficient is 1 when the determinant of the Jacobian is negative or null at the point *i* and 0 when $\det\mathbf{J_i}$ is positive. The procedure is then repeated for each distorted element and finally, the residual vector is derived from the summation over *p*, *p* being the index of all the elements in the mesh having the node *j* in their connectivity.

$$\mathbf{V}_j = \sum_p \sum_{i=1}^{n} \alpha_i \cdot \nabla(\det\mathbf{J_i})_j \tag{2}$$

where $\alpha_i = 1$ if $\det\mathbf{J_i} \leq 0$ and $\alpha_i = 0$ if $\det\mathbf{J_i} > 0$.

The modification of node locations is based on equation (3) where $X_j$ and $X'_j$ are the old and the new coordinates of the node *j*, and *w* is a factor depending on the scale of the structure, taken here as a percentage of the average edge length, *averLength*, taking into account the dimension of the mesh. The directional vector is finally normalized with the Euclidian norm so that $\mathbf{V}_j / \|\mathbf{V}_j\|$ has no dimension.

$$X'_j = X_j + \frac{\mathbf{V}_j}{\|\mathbf{V}_j\|} * w * averLength \tag{3}$$

In addition to the algorithm, maximal node displacements are constrained so that the regularized mesh still fits the patient morphology. The constraints for internal and external nodes differ but they are both based on a percentage of the displacement of the nodes from their initial positions, computed after the M-M algorithm (with a small percentage for external surface nodes in order to still fit the patient geometry).

3. Results

The regularization method is first evaluated on the simple test case presented in Figure 1-a. The cubic structure is meshed with hexahedrons starting from a controlled irregular mesh, shown in Figure 1-b. To get this distorted mesh, the node located inside the original cube was manually moved. The minimum Jacobian determinant is - 0.1125, thus no FE analysis is feasible. The regularization method succeeded, providing a regular mesh (figure 1-c) with a threshold value of $10^{-4}$ for the Jacobian determinant. Increasing the lower admissible value of the Jacobian determinant in the algorithm also improves the meshing and allows the convergence towards perfect cubic elements (figure 1-d).

The following clinical application concerns the automatic mesh generation of human faces in orthognatic surgery (Chabanas et al., 2003). The generic model used to run the M-M algorithm is plotted in Figure 2a and the regularized meshing of the new patient is plotted in Figure 2b. The human face mesh is made of 2884 elements and 4216 nodes and represents the soft tissues (skin, muscles and fat tissues) as a homogenous material. The

M-M algorithm generates 149 irregular elements that were detected by the procedure. As an example, a distorted element selected within the lips is plotted before and after regularization, in Figure 2c.

It took about one minute and 130 iterations (on a DEC Alpha 500 MHz computer) to correct the irregular elements. The new mesh remains very close to the one generated by the M-M algorithm and no geometrical difference can be visually observed. More quantitatively, among the 4216 nodes of the mesh, 614 were finally moved by the iterative regularization technique, with a 2.2 mm mean displacement value (maximum displacement = 2.692 mm; minimum displacement = 0.001 mm). In this test, the Jacobian determinant threshold value was $10^{-9}$.

The regularization method was successfully applied to six other patient FE models generated by the M-M algorithm. Two of the six regularized meshes are presented in Figure 3. Note that the mesh after regularization is still close to the CT data. Table 1 summarizes the regularization computation time, the number of irregular nodes, node displacements and the number of shifted nodes. For all the test cases, 5% to 10% of irregular elements have been detected and automatically regularized. Despite the obvious variation in geometries, good results were obtained and the computation time was less than four minutes.

Recently, the combination of the M-M algorithm and of the regularization phase has been applied to FE orbit meshes. As for the face, these two processes were required to generate a great number of meshes from a manually meshed orbit used as an atlas. This generic mesh is composed of 1375 elements and 6948 nodes and represents the soft tissues of the orbit, i.e. the fat tissues, the muscles and the optic nerve as a homogeneous poroelastic material. It has been developed to simulate orbital surgeries and more specially exophthalmia reduction (Luboz et al., 2004) in a computer assisted diagnosis framework.

Eleven patient-specific meshes were generated with the M-M algorithm. Each mesh had irregular elements: approximately 158 elements (with a standard deviation of 28). The regularization phase achieved to correct all of them by moving around 566 nodes (standard deviation: 99) with a mean displacement of 0.11 mm (standard deviation: 0.03). In this test, the Jacobian determinant was set to $10^{-1}$. Table 2 summarizes the regularization computation time, the number of irregular nodes, node displacements and the number of shifted nodes. All irregular elements were automatically regularized by our algorithm. Figure 4 plots two patient specific meshes thus generated and regularized. After a mean regularization time of about 3 min (standard deviation: 40 seconds), each mesh was corrected without any visible change in the geometry of the mesh surface.

## 4. Discussion and conclusion

In previous studies, it was demonstrated, for simple anatomical structures like the femora, that the M-M algorithm is efficient at automatically generating different patient meshes from an existing regular FE mesh. But some problems occurred when the geometry of the modelled anatomical structure became complex. In that case, the meshes automatically generated by the M-M algorithm were found irregular for a FE analysis. This paper introduced a new, fully automatic regularization procedure (based on the Jacobian determinant) that applies to these kind of irregular meshes. The procedure was illustrated in a simple test case (cubic mesh) and it was successfully evaluated for the regularization of seven FE meshes of the human face and eleven FE meshes of the orbit.

The regularization algorithm succeeds to automatically correct the irregular meshes generated by the M-M method. The patient meshes can then be used to carry out a Finite Element Analysis (in orthognatic surgery for the face model and in orbitopathy surgery for the orbit model).

Nevertheless, one must first notice that this regularization algorithm has been tested on a mesh that was originally manually designed. This means that the original elements of the generic mesh, matched to patient data with the M-M algorithm, were designed to be as regular as possible (with hexahedrons and wedges). In other words, the generated patient mesh was probably "less irregular" than a rough and unstructured tetrahedral mesh deformed by the M-M algorithm would have been.

Another limitation of the method is our inability to guarantee that the regularization algorithm will correct any irregular mesh. Indeed, due to its formulation, the iterative process of the algorithm tries to find a global solution, without any theoretical guarantee to converge. As can be seen on tables 1 and 2, some mesh regularizations need more iterations than other ones, but all of them finally converge to a stable solution.

In the next phase, we plan to deal with other clinical applications involving other geometrical FE models such as shoulder and liver. Another important perspective is to include quality criteria for the FE mesh into the iterative regularization algorithm (warping factor, parallel deviation, aspect ratio, edge angle, skew angle or twist angle).

**Acknowledgments**

Stéphane Lavallée is acknowledged for his contribution on the elastic registration algorithm used in this work.

**List of figures**

**Figure 1 -** Test case: cubic meshing. (a) perfect cubic mesh, (b) irregular mesh, (c) first regular mesh and (d) regular mesh with det**J** > 0.1.

**Figure 2** – (a) generic FE mesh of the face which leads to (b) a FE mesh of a patient face by applying the M-M algorithm. (c) example of the regularization procedure on a element.

**Figure 3** - Application of the M-M algorithm and the regularization phase to two patients with relatively different morphologies for the face. There is few visible difference between the real morphologies (top) reconstructed using the CT scan and the FE models obtained via the M-M algorithm coupled with the regularisation procedure.

**Figure 4** - Application of the M-M algorithm and the regularization phase to two patients with significant differences in orbit morphologies. The mesh at the left is the atlas that is deformed to fit the morphology of the other patients, thus creating patient-specific FE meshes.

**Table 1 -** Computational results for the regularization of the seven human face meshes.

**Table 2 -** Computational results for the regularization of the eleven orbit meshes.

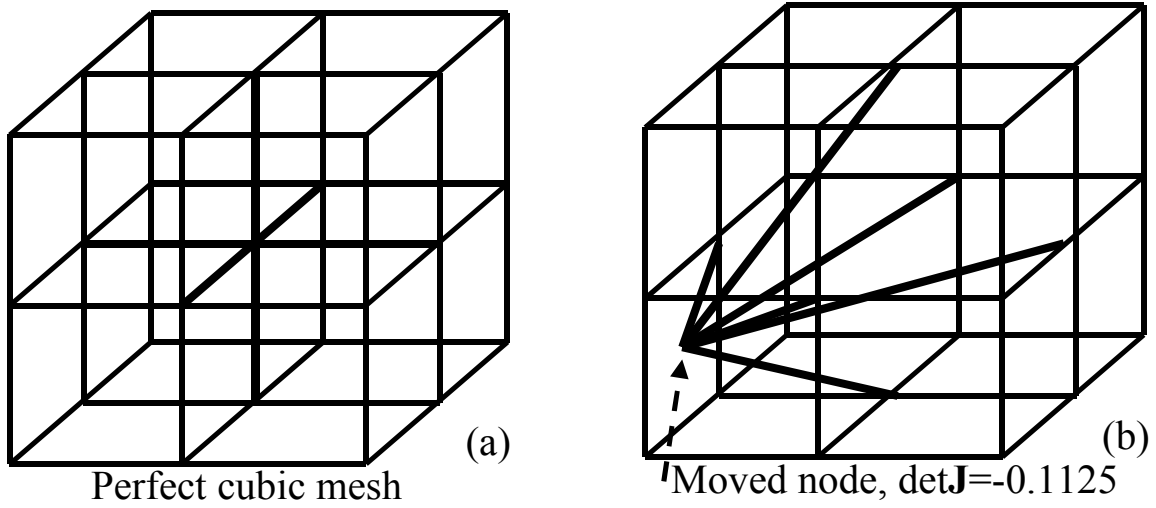

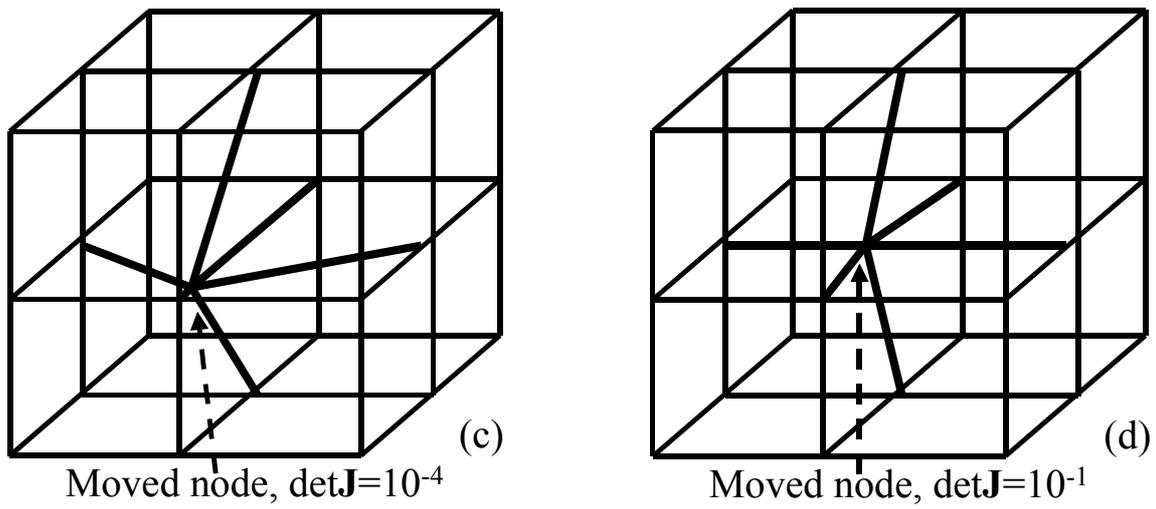

Figure 1

(a) Perfect cubic mesh
(b) Moved node, det**J**=-0.1125
(c) Moved node, det**J**=10⁻⁴
(d) Moved node, det**J**=10⁻¹

REGULARIZATION OF A MESH GENERATED WITH THE MESH-MATCHING ALGORITHM
Vincent Luboz, Pascal Swider, Yohan Payan

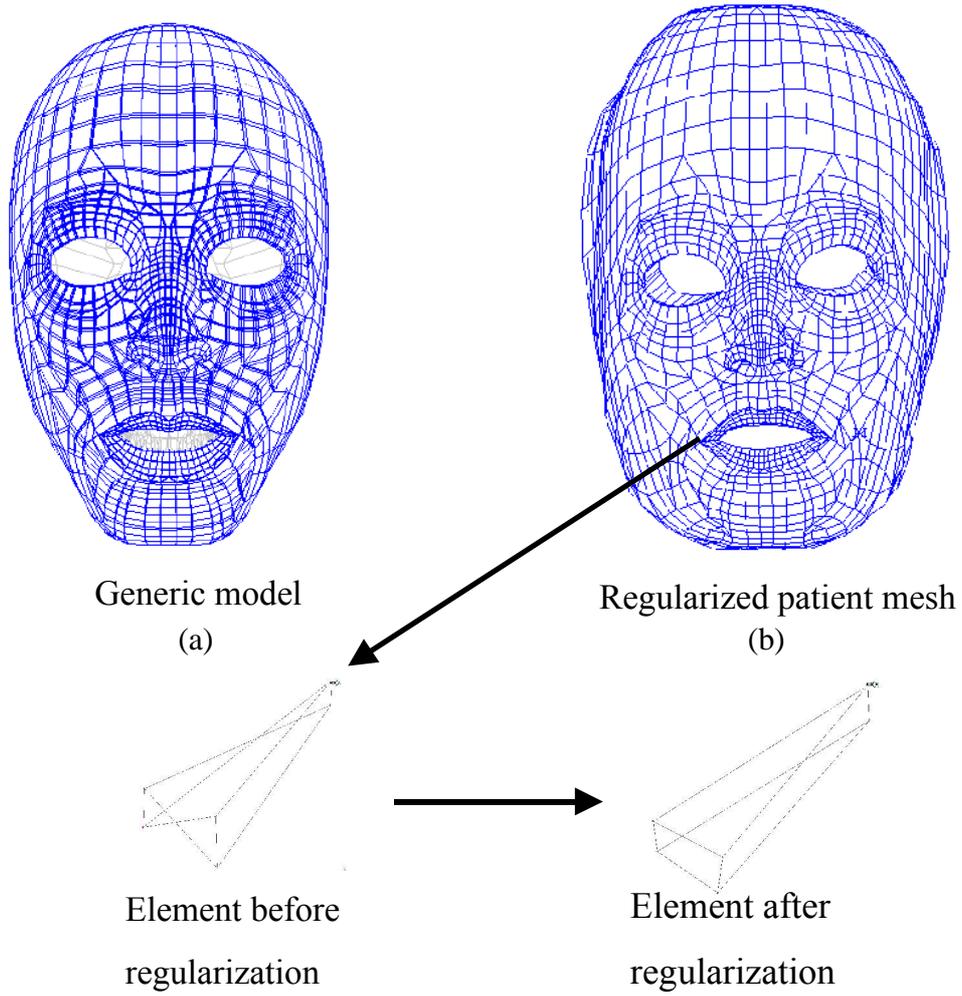

Generic model
(a)

Regularized patient mesh
(b)

Element before regularization

Element after regularization

(c)
Figure 2

REGULARIZATION OF A MESH GENERATED WITH THE MESH-MATCHING ALGORITHM
Vincent Luboz, Pascal Swider, Yohan Payan

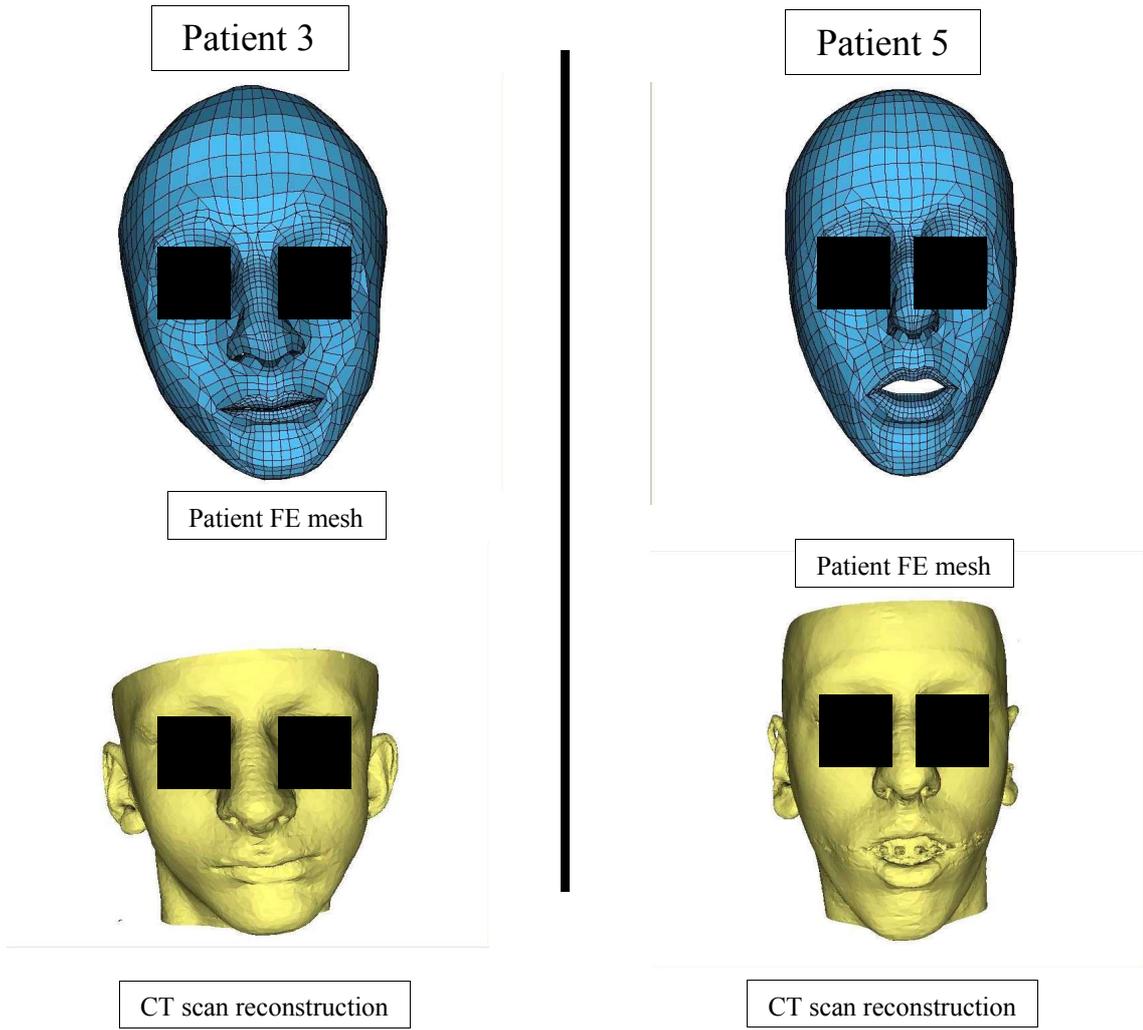

Figure 3

Regularization of a mesh generated with the Mesh-Matching algorithm
Vincent Luboz, Pascal Swider, Yohan Payan

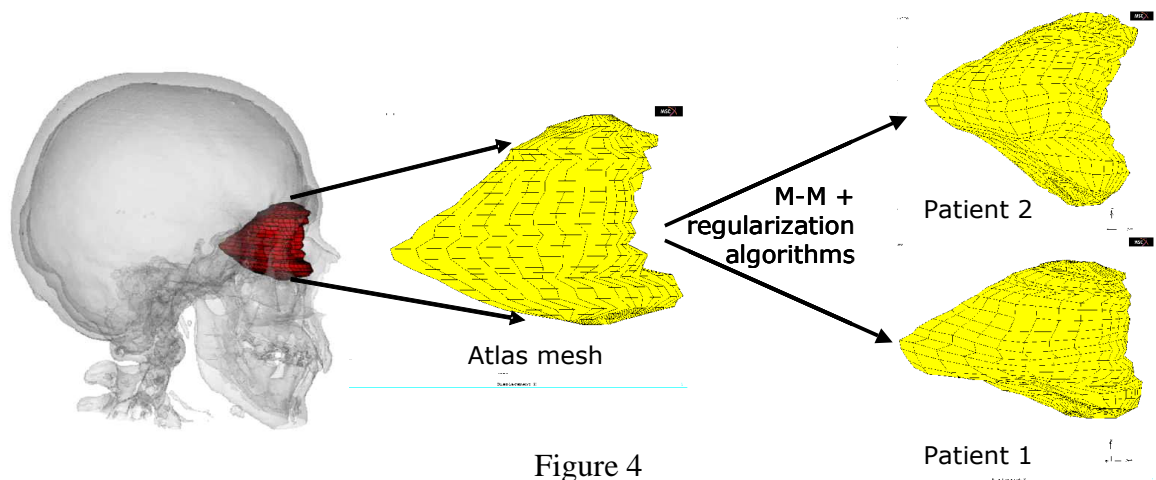

Figure 4

Regularization of a mesh generated with the Mesh-Matching algorithm
Vincent Luboz, Pascal Swider, Yohan Payan

**Table 1 -** Computational results for the regularization of the seven human face meshes.

|  | Number of irregular elements | Iterations number | Computation time | Min node disp. (mm) | Max node disp (mm) | Mean node disp. (mm) | Number of shifted nodes |
|---|---|---|---|---|---|---|---|
| Patient 1 | 149 | 130 | 1 minute | $10^{-3}$ | 2.69 | 0.22 | 614 |
| Patient 2 | 291 | 350 | 1 minute | $6.2\ 10^{-5}$ | 2.36 | 0.16 | 982 |
| Patient 3 | 268 | 300 | 1 minute | $2.3\ 10^{-5}$ | 3.36 | 0.21 | 1177 |
| Patient 4 | 191 | 450 | 3 minutes | $1.53\ 10^{-4}$ | 4.40 | 0.31 | 773 |
| Patient 5 | 234 | 350 | 4 minutes | $7.8\ 10^{-5}$ | 2.90 | 0.32 | 875 |
| Patient 6 | 253 | 350 | 3 minutes | $8.4\ 10^{-5}$ | 2.49 | 0.30 | 840 |
| Patient 7 | 239 | 350 | 3 minutes | $2.05\ 10^{-4}$ | 2.73 | 0.30 | 882 |

**Table 2 -** Computational results for the regularization of the eleven orbit meshes.

| | Number of irregular elements | Iterations number | Computation time | Min node disp. (mm) | Max node disp (mm) | Mean node disp. (mm) | Number of shifted nodes |
|---|---|---|---|---|---|---|---|
| Patient 1 | 276 | 400 | 5 minutes | $4.56\ 10^{-4}$ | 2.451 | 0.338 | 927 |
| Patient 2 | 202 | 200 | 3 minutes | $1.81\ 10^{-4}$ | 1.033 | 0.112 | 732 |
| Patient 3 | 203 | 100 | 1 minute | $1.26\ 10^{-4}$ | 1.21 | 0.115 | 798 |
| Patient 4 | 211 | 600 | 7 minutes | $1.07\ 10^{-4}$ | 1.175 | 0.101 | 660 |
| Patient 5 | 166 | 400 | 5 minutes | $2.88\ 10^{-4}$ | 1.135 | 0.103 | 728 |
| Patient 6 | 9 | 30 | 30 seconds | $0.03\ 10^{-4}$ | 0.41 | 0.004 | 39 |
| Patient 7 | 188 | 100 | 1 minute | $2.85\ 10^{-4}$ | 1.03 | 0.094 | 697 |
| Patient 8 | 11 | 30 | 30 seconds | $0.05\ 10^{-4}$ | 0.53 | 0.007 | 48 |
| Patient 9 | 232 | 200 | 3 minutes | $4.14\ 10^{-4}$ | 0.959 | 0.121 | 787 |
| Patient 10 | 237 | 300 | 4 minutes | $1.56\ 10^{-4}$ | 1.02 | 0.156 | 777 |
| Patient 11 | 8 | 30 | 30 seconds | $0.03\ 10^{-4}$ | 0.39 | 0.004 | 37 |